  \documentclass[aps,preprint,superscriptaddress]{revtex4-1}
\usepackage{graphicx} 
\usepackage{mathtools}
\usepackage{lineno}
\usepackage{lipsum}

\begin{document}
\title{A dusty plasma model for vortex structure in Jupiter's atmosphere }
\author{Modhuchandra Laishram}\email{modhu@ustc.edu.cn}
\address{CAS Key Laboratory of Geospace Environment and Department of Engineering 
and Applied Physics, University of Science and Technology of China, Hefei 230026,
China}
\author{Ping Zhu}\email{pzhu@ustc.edu.cn}
\address{CAS Key Laboratory of Geospace Environment and Department of Engineering 
and Applied Physics, University of Science and Technology of China, Hefei 230026,
China}
\address{KTX Laboratory and Department of Engineering and Applied Physics, 
University of Science and Technology of China, Hefei 230026, China}
\address{Department of Engineering Physics, University of Wisconsin-Madison, 
Madison, Wisconsin 53706, USA}

\author{Devendra Sharma}
\address{Institute for Plasma Research, HBNI, Gujarat, India, 382428}
 \date{\today}

 \begin{abstract}
Structural changes of self-organized vortices in 
Jupiter's atmosphere such as Great Red Spot (GRS) and White Ovals are 
demonstrated using an electrostatically bounded charged dust cloud in an 
unbounded streaming plasma as the prototype for various 
driven-dissipative complex flow systems in nature.  
Using a 2D hydrodynamic model, the 
steady state flow solutions are obtained for the volumetrically driven dust 
cloud in a bounded domain of aspect-ratio of 1.5 relevant 
to the current size of GRS and a driving sheared ion flow similar to the 
part of zonal jets streaming through the GRS. 
These nonlinear solutions reveal many similar characteristic 
features between the steadily driven dust circulation in laboratory experiments 
and the vortices in Jupiter's atmosphere. Starting from the continuous 
structural changes, the persistence of high-speed collar ring around 
the quiescent interior of uniform vorticity of GRS and White Ovals are 
interpreted as a consequence of changes in internal properties 
related to kinematic viscosity rather than the driving fields. 
This analysis also sheds light on the roles of driving field, boundaries, 
and dynamical parameters regime in determining the characteristic size, 
the strength, the circulating direction, 
and the drift of the vortices in Jupiter's atmosphere and other relevant 
driven-dissipative flow systems in nature.
\end{abstract}
\maketitle

\begin{figure}
\begin{center}
\includegraphics[width=0.99\textwidth]{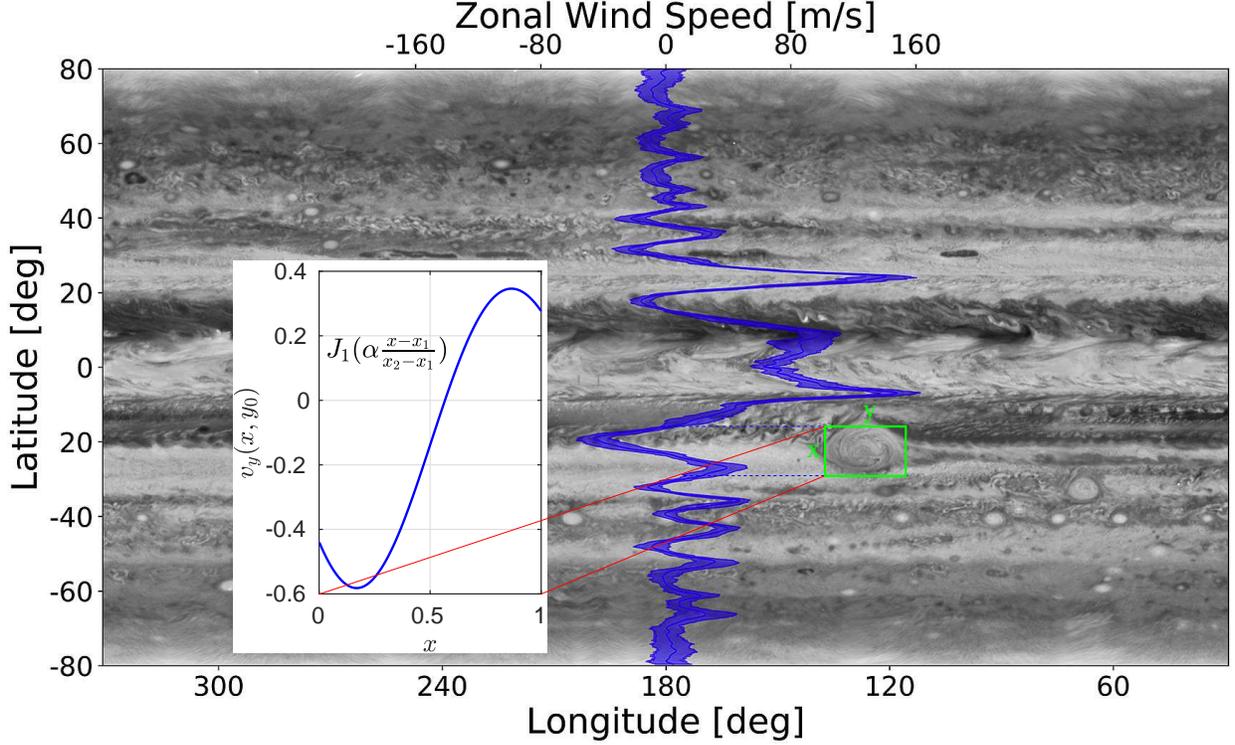}
\caption{Jupiter's zonal jets profile (inset) taken from the region in green box
 in the Hubble Space Telescope (HST) image of Jupiter. Reproduced with permission from 
{\it Planet. Space Sci.}~{\bf 155}, 2-11 (2018)~\cite{JOHNSON2018}.
\label{GRS_Jupiter}}
\end{center}
\end{figure}

Jupiter is the largest planet in solar system whose atmosphere 
supports all kinds of dynamics starting from small-scale instabilities 
and turbulence up to large-scale steady zonal jets and colorful 
vortices~\cite{Marcus,doi:10.1146/annurev.aa.31.090193.002515,VASAVADA1998265}. 
The most distinctive feature of Jupiter's atmosphere is its banded structure 
(Zones and Belts correlated with the sheared zonal jets) and immense vortices 
such as Great Red Spot (GRS) and White 
Ovals~\cite{JOHNSON2018,2041-8205-797-2-L31,CHOI2010359,Marcus771}. 
The GRS is a giant, long-lived, anti-cyclonic cloud vortex of the size $14.1^0 $ 
longitude by $9.4^0 $ latitude (where $1^0=1160~km$), 
situated at central latitude $22.3^0S$ of the Jupiter's 
atmosphere~\cite{1538-3881-155-4-151}, whereas the White Ovals are 
relatively small vortices observed at latitude $33.8^0S$, $41.8^0S$, 
and $19.0^0N$~\cite{2041-8205-797-2-L31,VASAVADA1998265,Marcus771,CHOI2010359}. 
Both GRS and White Ovals have quiescent core region of uniform vorticity 
surrounded by collar rings of high velocity
that dissociate the core region from the surrounding weak flows
~\cite{doi:10.1029/JA086iA10p08751,VASAVADA1998265}. 
Various spacecraft measurements have reported that GRS was first observed 
as a long pale hollow up to 1850s, then it became dusky elliptical rings 
up to 1870. Furthermore, it has been shrinking in size and accelerating 
in internal circulation ever since the late 1800s, while the streaming 
zonal jets have been globally stable around the 
GRS and White Ovals as shown in Fig.~\ref{GRS_Jupiter} 
~\cite{Roger1,SIMONMILLER2002249,JOHNSON2018}.   

Recently, NASA's spacecraft (JUNO-2018) has once again confirmed most
of the earlier observed characteristics of Jupiter's atmosphere, such
as the internal rotating period of GRS is decreasing, the giant
vortex is shrinking with rate $0.19^0$/year along latitude,
$0.048^0$/year along longitude, the overall structure is turning more
circular with a negligible change in the streaming zonal
jets~\cite{1538-3881-155-4-151,JOHNSON2018}. However, many puzzles
over the physical interpretation of Jupiter's vortices remain
unsolved even after many years of observations and analysis,
including the driving mechanism, 
the horizontal drift of GRS, the continuous structural
change, the presence of high-speed collar rings around the quiescent
interior, the long-life persistence of the vortices, and the actual
three-dimensional structure, among others~\cite{SIMONMILLER2002249,
  VASAVADA1998265,2041-8205-797-2-L31,Roger1}.
There are various single fluid models using the concept of potential
vorticity that interpret the vortices as Rossby soliton or turbulent
inverse cascade driven by Coriolis force of the
Jupiter~\cite{doi:10.1146/annurev.aa.31.090193.002515,doi:10.1063/1.166007}.
However, it has many criticisms regarding the flow profiles, the continuous
structural changes, and the horizontal drift velocity among many 
others~\cite{doi:10.1063/1.166007,doi:10.1175/1520-0469(1989)}.
Further, there are proposals that both the streaming zonal jets and
the vortices of Jupiter's atmosphere are driven by the moist convection
or the thermal convective instability from the deep interior, and the
energy from small-scale eddies, but these are doubtful to be the main
drivers of such giant vortices~\cite{Ingersoll}. The actual mechanisms
for the dynamics and peculiar characteristics features of these vortices still remain
unclear.

On the other side, the GRS and White Ovals are reported as
driven-dissipative complex flow systems consisting of multiple species
$NH_3$, $CH_4$, $Ar$, $NH_4SH$, and $H_2O$-ice in dynamic equilibrium
with the zonal jets and other in-out forcing factors such as Coriolis
effect and thermo-convection moist~\cite{LOEFFLER2016265,
  WEIDENSCHILLING1973465,doi:10.1146/annurev.aa.31.090193.002515}.
Moreover, the size, the strength, and the direction of vortices in
Jupiter's banded structure are strongly correlated with the position
and shear strength of the streaming jets, which we believe to be the
main driver for the Jupiter's
vortices~\cite{Roger1,1538-3881-155-4-151,2041-8205-797-2-L31,VASAVADA1998265}.
And surprisingly, we find the characteristics of GRS and 
White Ovals are found too much resemble that of bounded dust cloud circulation 
in an unbounded streaming plasma
~\cite{doi:10.1063/1.4929916,doi:10.1063/1.4916065,1367-2630-5-1-333,
PhysRevE.95.033204}. For example, in one of recent laboratory experiments 
(please see fig. S1 in the Supplementary Materials) and its theoretical 
formulation at a higher flow velocity 
regime~\cite{doi:10.1063/1.4929916,PhysRevE.95.033204}, we have observed 
compelling nonlinear features of dust-vortex dynamics having a nearly 
uniform vorticity in the core region surrounded by highly shear layers, and many 
regions of highly accelerating and deceleration in velocity field 
(please see fig. S2 in the Supplementary Materials), which are 
very similar to the flow characteristics of the Jupiter's GRS and White 
Ovals 
~\cite{doi:10.1063/1.4929916,PhysRevE.95.033204}. 
These common characteristics and physics shared by both systems motivates us to adopt 
the dynamics of the confined dust
cloud in the streaming plasma as a prototype for studying
various characteristics of GRS and White Ovals. Further, it has been
well known that GRS has thin, but wide upper layers along with a
relatively steady vertical stratified deeper layers inside, which does
not affect much on the dynamics of upper
layers~\cite{doi:10.1146/annurev.aa.31.090193.002515,VASAVADA1998265,Marcus771}.
Therefore, this structure allows us to approximate the surface dynamics
independent of the deep interior, which is closely similar to the
dynamics within the 2D cross-section of the electrostatically confined dust
cloud in the streaming plasma.

Considering the resemblance of dynamics of GRS and White Ovals 
to that of the bounded dust cloud in the streaming plasma which 
follows 
the incompressible 
and isothermal conditions, 
the dynamics of both systems in a 2D $XY$-plane, 
though vastly different in appearance, may be modeled using the 
modified Navier-Stokes equations in terms of the stream function 
$\psi\hat{z}$ and the flow vorticity $\omega\hat{z}$ as 
follows~\cite{landau2013fluid,PhysRevE.95.033204,doi:10.1063/1.5045772},
\begin{eqnarray} 
\nabla^2\psi&=&\omega,
\label{streamfunction-equation}\\
\frac{\partial {\omega}}{\partial t} +
({\bf u} \cdot \nabla) {\omega}&=&\mu \nabla^2 \omega
-\xi(\omega-\omega_s)-\nu\omega.  
\label{vorticity-equation}
\end{eqnarray}
Here, $\hat{z}$ is unit vector normal to the $XY$-plane, 
${\bf u}$ is the dust flow velocity, $\omega_{s}$ is the collective 
vorticity source from the sheared streaming background plasma. 
And the dynamic regime is determined by system parameters $\mu$, $\xi$, and 
$\nu$~\cite{PhysRevLett.68.313,doi:10.1063/1.4887003,PhysRevE.91.063110}.
For a laboratory glow discharge argon plasma, a typical set of parameters are
$n \simeq 10^{9}$ cm$^{-3}$, $T_{e}\simeq 3 eV$, $T_{i}\simeq 1 eV$, 
with system size $L_x \sim 10$ cm, and ions shear flow strength $U_{0}$ 
equivalent to the fraction of ion acoustic velocity $c_{s}=\sqrt{T_{e}/m_{i}}$, 
the value of parameters are $\xi\sim 10^{-4}~ U_{0}/L_x$, 
$\nu\sim 10^{-3}~ U_{0}/L_x$, and $\mu\sim 1\times 10^{-4} ~ U_{0}L_x$ 
respectively~\cite{PhysRevLett.88.065002,PhysRevE.91.063110,PhysRevE.95.033204}.
In the case of Jupiter's vortices, ${\omega}$ and ${\psi}$ are the 
relative vorticity and corresponding stream function of the clouds 
driven by shear zonal jets of vorticity $\omega_{s}$. 
And $\mu$ takes the role of kinematic viscosity of the driven clouds, 
$\xi$ is the interactions coefficients of the clouds with the streaming 
zonal jets, and $\nu$ is the interactions coefficients with the stationary 
background which maintain the steady flows. 
The corresponding absolute vorticity $\omega_{abs}$ in 
presence of stretching vorticity $\omega_{st}$ (3D effects) and 
Coriolis force at the planetographic latitude $\phi$ and angular velocity 
$\Omega$ of the planet is 
{$\omega_{abs} = \omega+\omega_{st}+2\Omega sin(\phi)$} 
~\cite{doi:10.1146/annurev.aa.31.090193.002515}. However, the present 
analysis emphasizes only on the $\omega$ and $\psi$ of the steady 
flow because the effect of stretching vorticity and Coriolis force are 
only to change the strength of absolute vorticity uniformly. 

The steady-state solutions of the above set of 
equations (\ref{streamfunction-equation})-(\ref{vorticity-equation}) 
are obtained using proper boundary conditions and vorticity sources 
relevant to the bounded Jupiter's vortices~\cite{Roger1,SIMONMILLER2002249,
PhysRevE.91.063110,doi:10.1063/1.5045772}.
Real confined systems allow having an arbitrarily shaped cross-section
determined by the confining potential.  However, for simplicity, we
select a rectangular domain of aspect-ratio $L_y/L_{x}=1.5$ relevant
to the current size of the GRS displayed in
Fig.~\ref{GRS_Jupiter}~\cite{JOHNSON2018,Roger1}. Within the domain,
the GRS is a partially bounded quasi-steady circulation in presence of
the stable streaming zonal jets, whose profile has a westward peak at
$19.5^0 S $ and a relatively weak eastward peak at $26.5^0 S$. This
may be the reason for the existence of horizontal drift of the GRS
towards the West~\cite{Roger1,1538-3881-155-4-151,JOHNSON2018}.  Thus,
the main driver $\omega_s$, i.e., the streaming ions in our model is
considered to take a profile similar to a portion of the sheared
zonal jets streaming through the GRS as highlighted (by blue color) in
the Fig.~\ref{GRS_Jupiter}.
\begin{figure*}[t]
\includegraphics[width=1.05\textwidth]{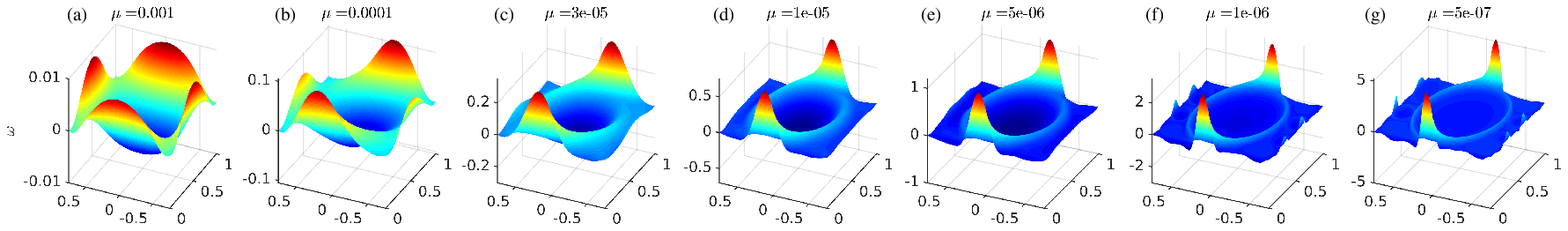}\\
\includegraphics[width=1.05\textwidth]{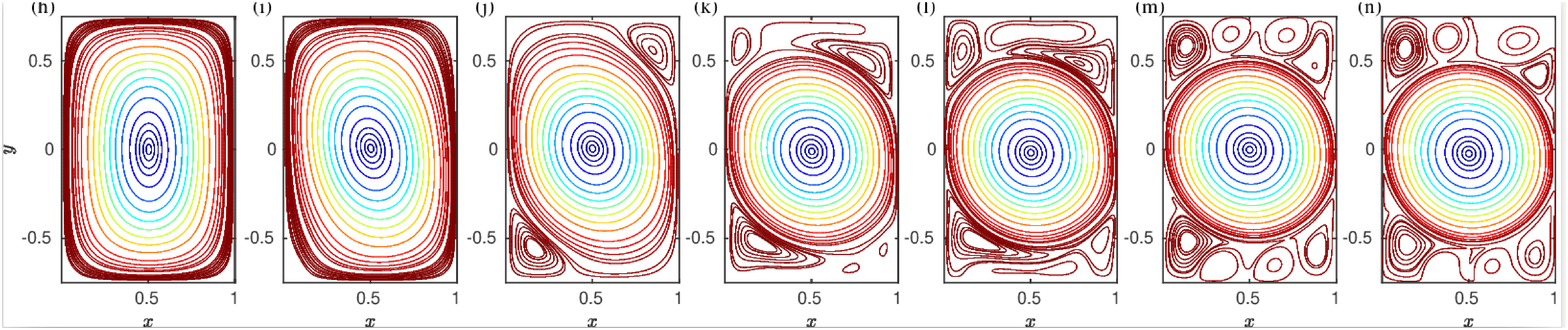}\\ 
\caption{Structural changes in steady-state dust vorticity (a) to (g), and 
corresponding streamline pattern (h) to (n), for a wide range of parameter 
$\mu$ from $ 1\times 10^{-3} ~ U_{0}L_x$ to $8\times 10^{-7} ~ U_{0}L_x$ 
and fixed other system parameters.
\label{solutions}}
\end{figure*}

Now, a series of the steady-state dust flow structure 
in term of the vorticity($\omega$) and corresponding streamline patterns 
(i.e., contours of $\psi$) in the rectangular $XY$-plane are displayed 
in Fig.~\ref{solutions}, for a wide range of kinematic viscosity $\mu$ 
and fixed other dependance on $\omega_s$, $\xi$, and $\nu$. 
In the case of highly viscous regime $\mu \sim 10^{-3} U_{0}L_x$, 
the flow vorticity along boundary is relatively weak, symmetric,
and uniformly diffusive throughout the whole domain as shown 
in Fig.~\ref{solutions}$(a)$. And the corresponding streamline pattern is 
the rectangular circulations following the geometry of confined domain 
as shown in Fig.~\ref{solutions}$(h)$. 
Then, decrease in $\mu$ leads 
the vorticity profile to strengthen and become asymmetric as shown in 
Fig.~\ref{solutions}$(b)-(c)$. The corresponding 
streamline in Fig.~\ref{solutions}$(i)-(j)$ turn into elliptical  
circulations that can retain more angular velocity $(\approx 2\omega$) 
without a significant change in the angular momentum of the system. 
Thus, the dynamical changes with the decrease in $\mu$ 
give a new state of the flow that retains more momentum or energy, 
and as a consequence, the new flow structure starts to form because of the 
dynamical regime rather than the geometry of bounded domain. 
Thus, for a further decrease in $\mu$, the asymmetry in 
vorticity profile gets enhanced such that highly sheared layers 
develop near the boundary along the driving ions and flatten near the 
boundary across the driver as shown in Fig.~\ref{solutions}$(d)-(e)$. 
The relative increase in convective transport enables the vorticity near 
the boundaries to convect along the streamlines and then dissipate to 
the background instead of diffuse directly towards the interior region. 
Therefore, the corresponding streamline patterns in 
Fig.~\ref{solutions}$(k)-(l)$ become more circular and turn into a new 
self-organized state with a circular core region surrounded by high-speed 
collar layers that dissociate the core from the surrounding regions 
filled with weak and elongated vortices. This qualitative change in vorticity 
or streamline patterns takes place through a critical parameter $\mu^*$, and 
this phenomenon is known as nonlinear structural 
bifurcation~\cite{PhysRevE.95.033204,GHIL2004149}. 
Further, in the case of highly nonlinear regime $\mu \le 10^{-6} U_{0}L$, 
the driven system retains more momentum and hence the vorticity along the 
collar layers and near the boundaries are strengthened, developing a flat 
uniform vorticity core region in the interior as shown in 
Fig.~\ref{solutions}$(f)-(g)$. The corresponding streamline patterns 
in Fig.~\ref{solutions}$(m)-(n)$, indicate that there is no significant 
change in the circular core region, however, the weak 
and elongated vortices near the boundaries also get strengthened and 
becomes more circular like the primary core vortex. 
\begin{figure}
\begin{center}
\includegraphics[width=0.99\textwidth]{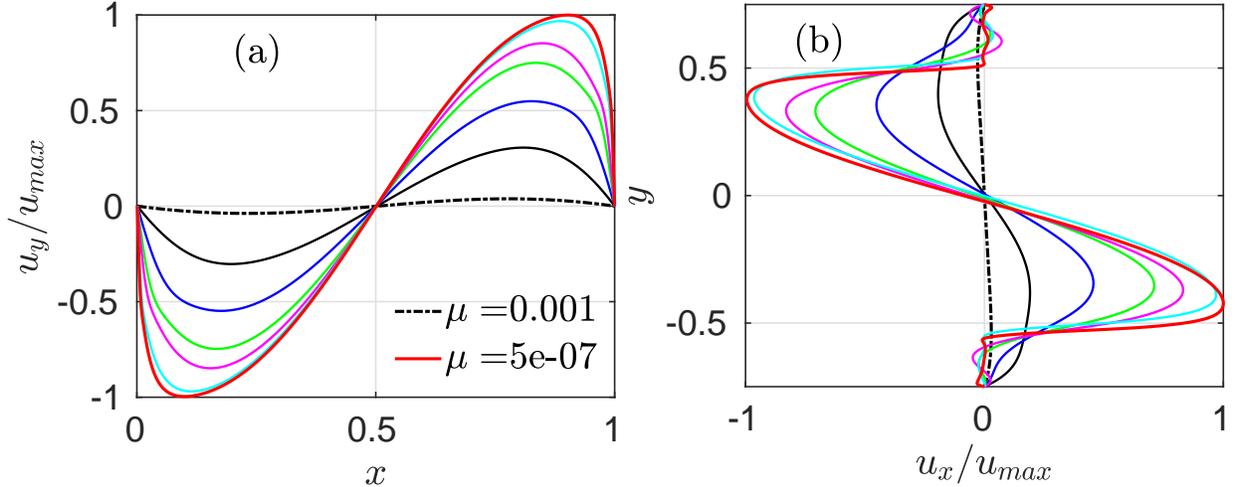}
\caption{Cross-section profiles of steady dust flow 
velocity (a) $u_y/u_{max} $ and (b) $u_x/u_{max}$ for the wide range of 
kinematic viscosity $\mu$. 
}
\label{driver}
\end{center}
\end{figure}

The structural changes in the vorticity and corresponding streamline
patterns of the bounded dust flow are again demonstrated in terms of
cross-section profiles of velocity $u_x$ and $u_y$ passing through the
center of the circulation $(x_0, y_0)$ as shown in
Fig.~\ref{driver}(a) and (b).  Here both $u_x$ and $u_y$ strengthen in
amplitude with decreasing $\mu$. And along the transitions, the
monotonic variation of $u_x$ along $y$-direction near boundary turns
oscillatory when the viscosity drops below the critical viscosity
$\mu^*$, indicating the nonlinear structural bifurcation and the
emergence of small-scale dynamics in the system. Further, with
increasing nonlinearity, both $u_x$ and $u_y$ approach constant
steady flow speeds with a single shear scale between two equidistant
opposite peaks, denoting the emergence of the high-speed collar layers
or virtual boundary (separatrix) which separates the circular core 
region of uniform vorticity away from the weak flows near boundaries. 
The characteristic size of the core region is determined by the
dominant scale such as the smallest distance between two opposite
boundaries or shear scale of the driving
field~\cite{PhysRevE.95.033204,PhysRevE.91.063110}.

The series of steady flow structures of the bounded driven dust cloud 
with changing $\mu$ in Fig.~\ref{solutions}
sheds light on the similar phenomena of structural 
changes of GRS and White Ovals, which are turning more circular without 
changing the driver zonal jets~\cite{Roger1,Barrado}. It 
indicates that the structural changes in vortices of the Jupiter's atmosphere 
are mainly due to a change in its internal properties related to the 
viscosity $\mu$ rather than the changes in the drivers. 
Therefore, the driven vortex is accelerating, 
the circulation period is decreasing, and the structure becomes more circular 
without a significant change in the total angular momentum of the whole system. 
Further, the emergence of 
the self-organized state with a circular core region of 
uniform vorticity surrounded by high-speed collar layers 
is quite similar to the presence of quiescent interior and 
high-speed collar rings of GRS’s and White 
Ovals~\cite{doi:10.1029/JA086iA10p08751,VASAVADA1998265}.  
The primary core region of uniform vorticity displayed 
in the Fig.~\ref{solutions} and the corresponding velocity 
profile in Fig.~\ref{driver}(a) and (b) at highly nonlinear regimes 
satisfy the Prandlt-Batchelor Theorem 
($\frac{\partial\omega}{\partial\psi}\approx 0$ 
means $\mu\oint \nabla^2 u\cdot d{\bf l}\approx0$) i.e., the core region 
is free from viscous dissipation (please see fig. S3 in the 
Supplementary Materials)~\cite{batchelor_1956,PhysRevE.95.033204}.
Once the fully circular structure of 
uniform vorticity emerges in any of the driven-dissipative systems, it 
persists or does not change its characteristics in a wide range of 
nonlinear regimes, whereas the surrounding weak vortices change accordingly. 
This is a possible explanation why the similar flow systems such as GRS and 
White Ovals have been persisting for a long time, even 
more than hundreds of years. It further gives the intuition that the 
quasi-steady circular vortices in the Jupiter's atmosphere may persist for a long 
life in future unless any dissipative mechanism is developed in the system.  
Furthermore, the shear nature of the dust velocity profile $u_x$ and $u_y$ 
at highly nonlinear regime displayed in Figs~\ref{driver}(a) and (b) is 
in close agreement with the observations in dusty plasma 
experiments~\cite{doi:10.1063/1.4929916} and 
that of the North-South and the East-West global velocity profile of 
White Ovals BA (please see fig. S4 in the 
Supplementary Materials)~\cite{CHOI2010359,Marcus771}.

Among the notable issues with this model, 
the scales of the dust dynamics are not identical to the real Jupiter's 
atmosphere and it can not able to interpret the 
turbulence behaviors at the center of the GRS which are expected to be 
3D-effects from the interior~\cite{Ingersoll}. However, vortices in the 
whole banded structure of Jupiter atmosphere have different 
size, strength, and direction depending on the shear nature of the zonal jets. 
These phenomena support the argument that the streaming zonal jets are 
the dominant driver of the vortices even though the vortices have additional effects of 
Coriolis force and 3D-effects which may actually 
strengthen or weaken the absolute vorticity. 
In short, 
this work has demonstrated various observed characteristic 
features of Jupiter's vortices  
are the consequence of changes 
in internal properties related to kinematic viscosity 
rather than the driving fields.\\

\section{Acknowledgments}
Author L.~Modhuchandra acknowledges late Prof. P. K. Kaw, 
Institute for Plasma Research, India, for the invaluable 
supports and encouragements. The research was supported by the State 
Administration of Foreign Experts Affairs - Foreign Talented Youth 
Introduction Plan Grant No. WQ2017ZGKX065, the National Magnetic 
Confinement Fusion Program of China under Grant Nos. 2014GB124002 and 2015GB101004. 
Author P. Zhu also acknowledges the support from U.S. DOE grant
Nos. DE-FG02-86ER53218 and DE-FC02-08ER54975.  
The work used the resources of Supercomputing Center of University 
of Science and Technology of China. 
\end{document}